\documentclass[aps,prl,twocolumn,groupedaddress,superscriptaddress,showpacs]{revtex4-1}
\usepackage[hidelinks,colorlinks=true,linkcolor=blue,citecolor=blue,anchorcolor=blue,filecolor=blue,menucolor=blue,runcolor=blue,urlcolor=blue]{hyperref}
\usepackage{bm}
\usepackage{color}
\usepackage{graphicx}
\usepackage{amssymb}
\usepackage{amsmath}
\usepackage{ulem}
\usepackage{gensymb}
\usepackage{lineno}
\makeatother

\begin{document}
\title{Gravity-induced accelerating expansion of excited-state Bose-Einstein condensate}
\author{Lijia Jiang}
\affiliation{Institute of Modern Physics, Northwest University, 710127 Xi'an, China}
\author{Jun-Hui Zheng}
\email{junhui.zheng@nwu.edu.cn}
\affiliation{Institute of Modern Physics, Northwest University, 710127 Xi'an, China}
\affiliation{Peng Huanwu Center for Fundamental Theory, 710127 Xi'an, China}


\begin{abstract}
{The Bose-Einstein condensate (BEC) of excited states, provides a different platform to explore the interplay between gravity and quantum physics. In this Letter, we study the response of excited-state BECs to an external gravitational field and their dynamics under gravity when space is expanding. We reveal the anomalous response of the center-of-mass of the BEC to the gravitational field and the exotic gravity-induced accelerating expansion phenomena. We demonstrate that these effects result from the interplay among gravity, space and quantum effects. We also propose related experiments to observe these anomalies.}
\end{abstract}
\maketitle

The study of quantum matter in gravitational fields or in curved space-time continues to reveal fascinating phenomena, leading to novel physical concepts such as the Unruh effect, Hawking radiation, black hole entropy, etc \cite{Hawking1974, Unruh1976, Wald2001, Unruh2017}. These discoveries greatly advance our understanding of the nature of the vacuum and the thermodynamics of black holes \cite{Dodelson2008}. Meanwhile, with the development of laser cooling technology, the Bose-Einstein condensate (BEC) in ultracold atomic systems has become a highly clean and controllable platform for quantum simulations. Since the geometry of BECs is highly sensitive to gravity, they have also been used to measure gravitational field strength and to test the universality of free fall and the Einstein equivalence principle, facilitating the development of precise measurements \cite{Cronin2009, Bonse1983, Mashhoon1988, Greenberger1983, Geiger2011, Rosi2014, Xu2019, Staudenmann1980, Obukhov2001, Zych2018, Fray2004, Okon2011, Schlippert2014, Zoest2010,Aguilera2014}. In addition, with the realization of BECs in a microgravity environment \cite{Zoest2010, Lundblad2023}, the discovery of topological phase transitions of BECs in curved space \cite{ Aveline2020, Tononi2019, Tononi2020, Tononi2022, Carollo2022}, and the successful simulation of black holes and cosmic expansion by BECs \cite{Eckel2018, Banik2022, Bhardwaj2021}, more and more macrocosmic phenomena can be demonstrated in laboratories \cite{Viermann2022, Tian2022, Moreno2022, Torres2022}.

Therefore, BEC systems are promising objects for further studies of the interplay among quantum matter, gravity, and the geometry of space. Nowadays, in addition to ground-state BECs, excited-state BECs can also be prepared by different techniques in ultracold atom experiments, providing us novel opportunities to study unconventional systems out of thermal equilibrium, such as negative-temperature systems and soliton systems \cite{Rapp2010, Wirth2011,Becker2008, Denschlag2000}. So far, it is clear how gravitation affects the geometry of the ground-state BEC during its formation \cite{Carollo2022}. Moreover, the expansion redshift and Hubble friction of phononic excitation, spontaneous particle creation, the generation of bulk topological excitations, and other relevant dynamics during the inflation of ground-state BEC have been well studied \cite{Eckel2018, Llorente2019, Eckel2021, Banik2022, Bhardwaj2021, Viermann2022, Tian2022, Moreno2022, Torres2022}. Yet, the effects of gravity and spatial expansion on the excited-state BECs remain unclear and deserve systematic investigations.

In this Letter, we introduce a gravitational field to an excited-state BEC in static or expanding space and reveal the exotic gravitational effects. First, we demonstrate in a one-dimensional (1D) box potential well that, unlike the ground-state BEC, the center of mass of the excited-state BEC is lift by an external gravitational field,  exhibiting a ``repulsive" effect. Then, we reveal the phenomenon of gravity-induced accelerating expansion for the excited-state BEC in an expanding space, which is counter to our former impression that gravity slows down the expansion of matter. We explain the mechanism of these anomalies and propose currently achievable experiments to observe these phenomena.

We start with the simplest 1D case with a box potential well which can be realized using optical beams in experiments \cite{Gaunt2013}. The BEC confined in the well is described by a single macroscopic wavefunction $\Phi(x,t)$. Its dynamics under a gravitational field ${\bf G} = -g {\bf e}_x$ follows the Gross-Pitaevskii (GP) equation,
\begin{equation}\label{evoleq}
i\hbar \frac{\partial \Phi}{\partial t} =-\frac{\hbar^2}{2 M} \frac{\partial^2 \Phi}{\partial x^2} + \left[V(x)+M g x\right]\Phi + \lambda |\Phi|^2 \Phi,
\end{equation}
where the last term is the two-body repulsive interaction with $\lambda>0$ and $\Phi$ is normalized to one. The potential $V(x)=0$ for $|x|<L/2$ and is infinite elsewhere, where $L$ is the width of the well.

We first consider the case when the gravitational field is absent. If there is no trap, the GP equation is translation invariant and can be analytically solved. The self-consistent solution with a static dark soliton or a moving grey one that maintains its shape during propagation has been found \cite{Tsuzuki1971, Zakharov1974}.
In a trap, we numerically obtain the stationary self-consistent solutions $\phi_n(x,t) = \phi_n(x) e^{-i \epsilon_n t}$ of the nonlinear GP equation by cyclic iteration, where $n$ is the number of the wavefunction's nodes. The nodes indicate where the density of the BEC vanishes. When $\lambda/L \gg \hbar^2 k_n^2/2 M$, where $k_n = (n+1) \pi/L$ is the wavevector of the $n$-node state, the interaction dominates, and each node represents a center of a dark soliton with a size (which is about twice of healing length) much less than the wavelength $2\pi/k_n$. These solitons possess extremely long lifetimes ($\sim 10 s$) in experiments \cite{Becker2008}. As the increase of kinetic energy, the number of nodes also increases and when $\lambda/L \ll \hbar^2 k_n^2/2 M$, the kinetic energy becomes dominant. Then, the $n$-node state is similar to a free particle wavefunction and thus becomes sine-like in the box well. In this case, the condensate should also have a long lifetime, as the collision effects between atoms are relatively weak.

\begin{figure}[tbp]
    \centering
    \includegraphics[width=\columnwidth]{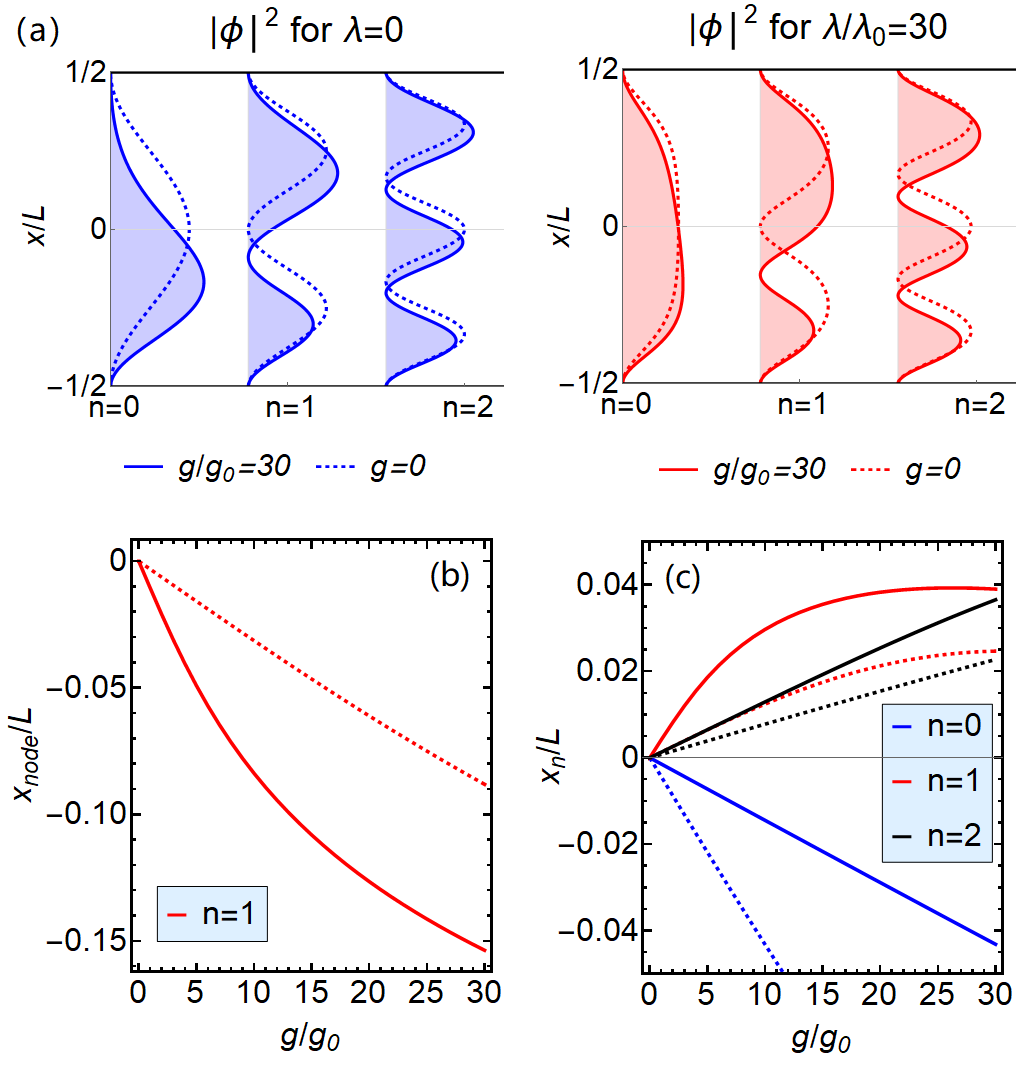}
    \caption{(a) The stationary density distribution $|\phi_n|^2$  ($n=0,1,2$) in a box potential well for $g=0$ and $30 g_0$, respectively. (b) The node position as function of field strength $g$ for $n=1$. (c) The center of mass ($x_n$) of BEC as a function of $g$. The units for the field strength $g$ and the interaction strength $\lambda$ are $g_0=\hbar^2/M^2 L^3$ and $ \lambda_0=\hbar^2/M L$, respectively. In subfigures (b) and (c), dashed lines are results at the noninteracting limit with $\lambda=0$ and solid  lines are results for the interaction case with $\lambda = 30 \lambda_0$.}\label{fig1}
\end{figure}

Hereafter, we focus on the effect of a gravitational field on the BEC. In Fig.\,\ref{fig1}(a), we plot the stationary density distribution $|\phi_n|^2$ and highlight the influence of gravity ($g/g_0=30$) for BEC with $\lambda/\lambda_0 = 0$ and $30$, separately, where $g_0 = \hbar^2/M^2 L^3$ and $\lambda_0 = \hbar^2/M L$ are the units. We find that the node positions for $g \neq 0$ are generally lower than the corresponding ones for $g=0$. Fig.\,\ref{fig1}(b) further shows how the node position decreases with the increase of $g$ for the 1-node state. In the kinetic-energy-dominant regime (see the case of $\lambda=0$ or the case with $n=2$ and $\lambda=30$), the downward movement of nodes under gravity implies
the gravitational redshift effect of matter waves \cite{Geiger2011, Greenberger1983, Bonse1983, Mashhoon1988}. For the interaction-dominant regime (see the case with $n=1$ and $\lambda = 30$), the downward movement of nodes under gravity shows that the soliton is like a normal particle in a gravitational field. This is consistent with the fact that the soliton has a negative inertial mass and is effectively subjected to an upward gravitational pull due to the lack of Thomas-Fermi density \cite{Becker2008}.

In Fig.\,\ref{fig1}(c), we plot the center-of-mass of the BEC, $x_n \equiv \langle \phi_n | \hat {x} |\phi_n \rangle$, with respect to the variation of the gravitational field strength $g$. Unlike the ground state, the position of the center-of-mass for the condensate of excited states increases with $g$. Dynamically, assuming that $g$ slowly increases with time, the curve $x_n(g(t))$ also depicts the adiabatic trajectory of the center-of-mass of the BEC (`adiabatic' means that the BEC stays at an instantaneous stationary state of the system with the same energy level). We define {\it gravitic susceptibility} for the adiabatic process as
\begin{equation}
    \chi_n = \frac{\delta x_n}{\delta g} = \frac{\delta \langle\phi_n |\hat x | \phi_n\rangle}{\delta g},
\end{equation}
which characterizes the response of the center-of-mass to the variation of the field strength. In the relatively weak gravity regime, both the results in Fig.\,\ref{fig1}(c) and numerical calculations for higher $n$-node states show that the gravitic susceptibility of the condensate of excited states is generally positive, which indicates an effective `repulsive' effect from gravity.
In addition, weak or moderate interaction does not qualitatively change the response of the BEC to gravity. In particular, the interaction effects become negligible gradually with the increase of $n$.

In the following, we provide an explanation on the above anomalous responses at the noninteracting limit where $\phi_n$ with different $n$ become orthogonal. The corresponding gravitic susceptibility for the $n$-node state in the adiabatic process is,
\begin{equation}
\chi_n  = 2 M \sum_{l\neq n} \frac{|\langle \phi_l | x |\phi_n\rangle|^2}{\epsilon_n-\epsilon_l}.
\end{equation}
For the ground-state BEC, $\chi_n$ is always negative since $\epsilon_0<\epsilon_n$ for all $n\geq 1$, which is consistent with our intuition on the gravitic effects. Meanwhile, the arithmetic mean of $x_n$ over the complete basis vanishes for a finite-size system, $\overline{x_n} \equiv \text{Tr} [\hat x] /\text{Tr}[\hat 1] = \int_{-L/2}^{L/2} x dx\big/\int_{-L/2}^{L/2} dx = 0$. Accordingly, the total susceptibility $\sum_{n} \chi_n$ is zero as well. Therefore, the gravitic susceptibility $\chi_n$ must be positive for some $n>0$ states.

In the adiabatic process of increasing $g$, the total work done by gravity on the $n$-node state is $W_n = - M \int g dx_n$. From Fig.\ref{fig1}(c), we can see that this work is positive only for $n=0$. Denoting the change of kinetic energy as $\delta E_k$ and that of the wavefunction as $\delta\phi_n$, we have
\begin{eqnarray}
\delta E_k - \delta W_n &=& \big[\langle \delta\phi_n|\frac{\hat p^2}{2M} | \phi_n \rangle + M g \langle \delta\phi_n|\hat{x}| \phi_n \rangle\big] +c.c \notag\\
&=& \epsilon_n \big[\langle \delta\phi_n| \phi_n \rangle +c.c\big] =0.
\end{eqnarray}
Hence, the work done by gravity completely transfers to the change of kinetic energy. In momentum space, we find that the particle's probability density distribution of the 1-node state is enhanced in the low-momentum regime and suppressed in the high-momentum regime by increasing $g$, totally opposite to the gravity effect on the ground state. Analogizing the kinetic energy with the internal energy in a thermodynamic equilibrium system, the increasing of gravitational potential energy is realized via releasing the BEC's internal energy, or in other words, by `cooling' the condensate itself.

For a finite $\lambda$, the wavefunctions $\{\phi_n|n=0,1,\cdots\}$ are not orthogonal anymore, and the work done by gravity will partially transfer to interacting energy. However, the numerical results show that the properties of stationary states are qualitatively in accord with the results in the $\lambda =0$ case, except for the interaction-induced flattening of the wavefunction (see Fig.\,\ref{fig1}(a)).

\begin{figure}[tbp]
    \centering
    \includegraphics[width=\columnwidth]{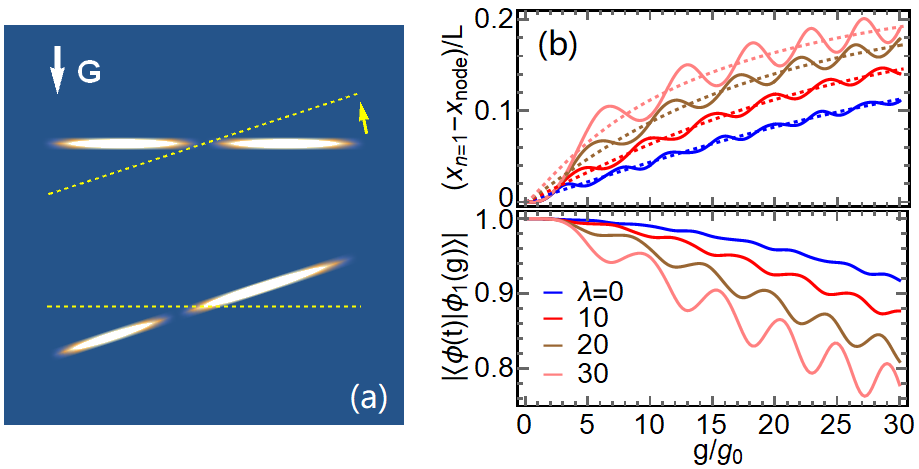}
    \caption{(a) Sketch for the mechanism of changing the gravitational field strength $g$ by adjusting the orientation of the tube-shaped BEC and (b) the normalized relative shift between the center of mass of the BEC and the position of the node, $\gamma \equiv (x_{n}-x_{\text{node}})/L$, with respect to $g$. Here, $n=1$. The dashed lines are for the adiabatic limit ($\dot{g}(t)\rightarrow 0$). The solid lines are the results of the dynamical real-time evolution for $g(t) = 10 t$ with $t \in [0, 3t_0]$, which qualitatively agrees with the adiabatic result except the oscillations. The unit for time is $t_0 = M L^2/\hbar$.
    }\label{fig2}
\end{figure}

Experimentally, a $n$-node BEC can be realized by using the phase imprinting technique \cite{Denschlag2000, Becker2008, Fritsch2020} or produced as defects in quench dynamics with a phase transition as a result of the Kibble-Zurek mechanism \cite{Lamporesi2013, Navon2015, Halperin2020, Kibble1976}. The field strength $g$ can be tuned by adjusting the angle between the tube-shaped BEC and the vertical gravitational field as sketched in Fig.\,\ref{fig2}(a). Considering the BEC of ${}^{87}\text{Rb}$ atoms in a box potential well with $L=10\mu m$, the units for the field strength and interaction strength are $g_0 =\hbar^2/M^2 L^3 =5.3\times 10^{-4} m/s^2$ and $\lambda_0 =\hbar^2/M L =7.7\times 10^{-39} kg ~ m^3/s^2$, respectively. Increasing $g$ from $0$ to $30 g_0$ can be realized by counterclockwise rotating the horizontally-placed BEC on the earth about $0.09^\circ$. If the well width is $L=5\mu m$, the rotating angle becomes $0.74^\circ$.

With the above setup, we numerically simulate the dynamical evolution of the BEC in a box potential well with time-dependent $g(t)=10t\cdot(g_0/t_0)$, where $t_0 = M L^2/\hbar$ (it equals to  $0.137 s$ for $L=10\mu m$). Since the center-of-mass $x_{n}$ and the node position $x_{\text{node}}$ move in opposite directions as $g$ increases, the normalized relative shift $\gamma \equiv (x_{n}-x_{\text{node}})/L$ can be easily identified in experiments by using the absorption imaging technique. In the upper panel of Fig.\,\ref{fig2}(b), we plot $\gamma $ with respect to $g$ for the case of $n=1$. In the adiabatic approximation (dashed lines), $\gamma $ grows  steadily as the increase of $g$. The real-time dynamical evolution of $\gamma $ (solid lines) shows clear oscillations around the adiabatic results, which represents the non-adiabatic effects from the tunneling between different instantaneous stationary states. Quantitatively, interaction significantly enhances both the magnitude of $\gamma$ and the amplitude of oscillation. The lower panel of Fig.\,\ref{fig2}(b) presents the overlap of the dynamical $\phi(t)$ and the adiabatic wavefunction $\phi_1(g)$, which indicates that the adiabatic approximation is easier violated for a larger interaction.
We emphasize that as the BEC is sensitive to the gravitational field ($\gamma$ is about $0.1\sim0.2$ for $g=30g_0$), this experiment set can, in turn, serve as a ruler to measure the gravitational field strength.

\begin{figure}[tbp]
    \centering
    \includegraphics[width=\columnwidth]{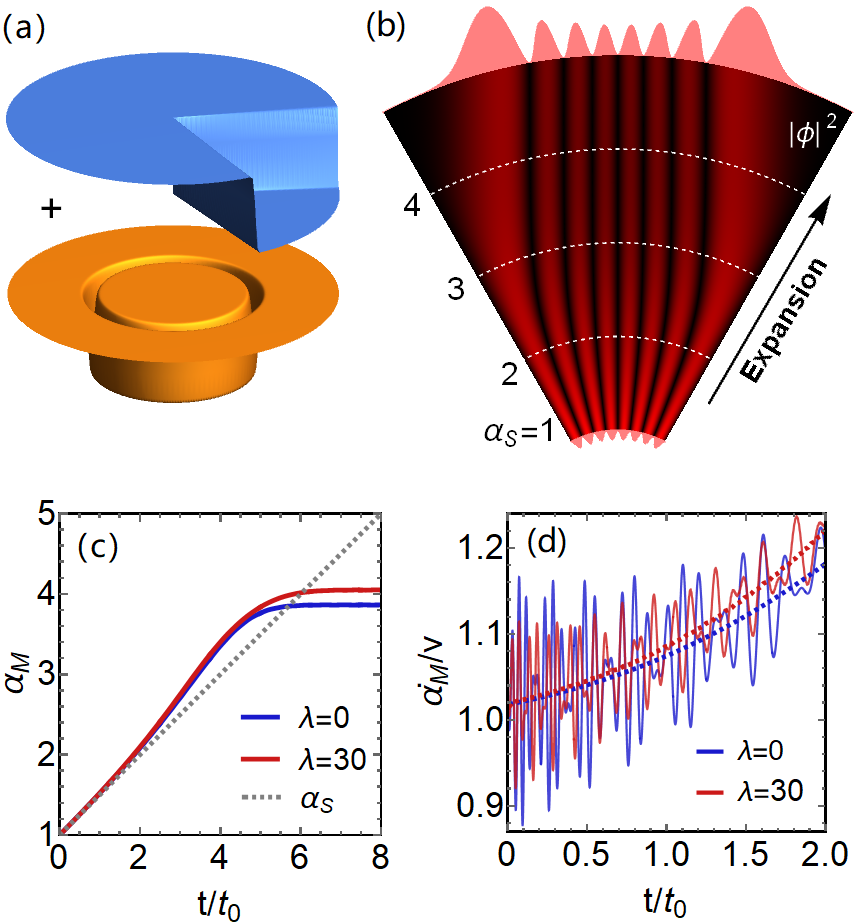}
    \caption{(a) Realizing box well in a ring potential. The initial BEC is supposed to be loaded in a sector of a ring trap with $\alpha_S=1$ and evolves as the enlargement of the radius of the ring. (b-d) The dynamics of BEC with $n=7$ during the expansion of space with the expansion rate $v =1/2t_0$, in the presence of the gravitational field $g=50 g_0$. (b) The dynamical particle number density distribution $|\Phi_n|^2$ for $\lambda=30$. With the expansion of space,  the matter wave redshifts and the BEC gradually accumulates at the boundary of the arc. At large $a_S$, the solitonic BEC ceases expansion and the BEC is fully confined by the central gravitational potential. (c) The dynamical expansion factor of the BEC with respective to time for $\lambda=0$ and $30$. (d) The ratio $\dot{\alpha}_M/v$ for the expansion rates of the BEC and space with respect to time $t$. The solid lines are dynamic simulation results, dashed lines are the results of adiabatic approximation.
    }\label{fig3}
\end{figure}

With the `repulsive' effect of gravity on the BEC of excited states, a natural question is how the BEC behaves in an expanding space when gravity is present. Below, we propose an experimental setup to search for the answer. By loading BEC in a ring trap, an expanding space can be simulated by tuning the ring's radius $r$ in experiments, as was done in Refs.\,\cite{Banik2022,Eckel2018,Bhardwaj2021,Carollo2022}. One needs to introduce an additional potential to confine the BEC to a sector of the ring trap so that the BEC has boundaries and a center (see Fig.\ref{fig3}(a) for the sketch). The BEC is set to locate in an effective central gravitational potential $Mg r|\theta|$, where $\theta$ is the angle in radians. This linear potential could be simulated by the laser beams. Then, by using the phase imprinting technique, the initial BEC of an excited state can be prepared.

The GP equation for such a setup reads:
\begin{equation}\label{evoleq2}
i\hbar \frac{\partial \Phi}{\partial t} =-\frac{\hbar^2}{2 M} \frac{\partial^2 \Phi}{r^2\partial \theta^2} +M g r |\theta| \Phi + \lambda |\Phi|^2 \Phi,
\end{equation}
where $\theta \in [-1/2,1/2]$ so that the gravitational field is centripetal. The spatial expansion is depicted by $r(t) = \alpha_S(t) L$, where the scale factor of space $\alpha_S(t) = 1 + vt$ changes with time uniformly \cite{Dodelson2008}. Without loss of generality, we select $n=7$ to visualize the dynamics of the BEC during the expansion. Fig.\,\ref{fig3}(b) presents the real-time density distribution in the ring for $\lambda=30$, $v =1/2t_0$, and $g=50$. At the beginning, the BEC co-moves with space during the expansion, thus the wavelength of matter waves increases, which reflects the expansion-induced redshift \cite{Eckel2018,Bhardwaj2021}. Moreover, the BEC gradually accumulates in the peripheral region of the sector, which implies that the BEC expands faster than space even under the attractive gravity. As the continuous expansion of space, the kinetic energy decreases and the central gravitational potential becomes comparable. The BEC finally ceases its expansion and becomes a bound state of the gravitational potential, and the wavelength does not further increase as well.

We introduce the effective radius $R(t) \equiv r \langle \phi| \text{Abs}[ \hat{\theta}] |\phi\rangle$ and matter expansion factor $\alpha_M(t) \equiv R(t) / R(t=0)$ to characterize the varying size of the BEC.  Fig.\,\ref{fig3}(c) presents the numerical results of $\alpha_M(t)$. The expansion of the matter is faster than that of space at the early stage ($\alpha_M > \alpha_S$) and ceases when $\alpha_S$ becomes sufficiently large ($\alpha_S\sim 4 $). The final size and the corresponding expansion factor of the BEC are larger for a stronger repulsive interaction, a weaker gravitational confinement, or a higher excited state $n$. Note that with current parameter set, the dynamical results are almost the same as the adiabatic result, and the overlap between the dynamical and instantaneous stationary wavefunctions is higher than $99 \%$.

In Fig.\,\ref{fig3}(d), we plot the ratio ($\dot\alpha_M/v$) of expansion rates for the BEC and space (solid lines). Likewise, non-adiabatic effects during the evolution bring oscillations around the expectation values from the adiabatic approximation (dashed lines). In the early stage of the expansion, the ratio on average increases faster and faster with time,  which means that the expansion of the BEC is not only faster but also accelerated. This indeed again reflects the effective `repulsive' effect of the gravitational field.

From another perspective, if we redefine time as $d\tilde t = dt/\alpha^2_S$ and the wavefunction $\tilde\Phi = \sqrt{\alpha_S}\Phi$, Eq.\,\eqref{evoleq2} can be rewritten as:
\begin{equation}\label{evoleq3}
i\hbar \frac{\partial \tilde\Phi}{\partial \tilde t} =-\frac{\hbar^2}{2 m} \frac{\partial^2 \tilde\Phi}{L^2\partial\theta^2} +m \tilde g L |\theta| \tilde\Phi + \tilde\lambda(t) |\tilde\Phi|^2 \tilde\Phi,
\end{equation}
where $\tilde g(t) = \alpha^3_S g $ and $\tilde \lambda(t) = \alpha_S \lambda$, and then the expansion process is equivalent to the dynamical evolution of the BEC with increasing field strength $\tilde g$ and interaction strength $\tilde \lambda (t)$ in a static space. Thus, the accelerating expansion of the BEC is consistent with the anomalous response of the condensate of excited states to gravity in static space.

In conclusion, we firstly reveal the anomalous gravitational response of excited-state BECs resulting from the finite-size effects and the orthogonality of quantum states. 
The response obeys the law of energy conservation --- the BEC increases its gravitational potential energy by `cooling' itself. Note that a stationary system in a gravitational field is equivalent to an accelerated moving system in the weightless environment ({\it the equivalence principle}) \cite{Okon2011}. Thus when treating the density node as a bubble, the anomalous response implies anomalous inertia behaviors of the system, compared to an accelerating (confined) classical liquid with bubbles. We also propose to observe the anomalous response by measuring the relative shift between the center of mass of the BEC and the node position, and to tune the gravitational field strength by slightly rotating an initially horizontally-placed tube-shaped BEC. The shift can be $10\%$-$20\%$ of the size of the BEC.

Secondly, we demonstrate that when the finite-sized space expands uniformly with time, during the early stages, the expansion rate of the excited-state BEC is accelerated by the attractive force of gravity and thus expands at a faster rate than the space. The resulting accumulation of the BEC in the peripheral region of the trap can be observed through real-time absorption imaging technique \cite{Anderson1995,Sherson2010}. This phenomenon of gravity-induced accelerating expansion significantly deviates from the impression of gravity in classical physics. We further demonstrate the equivalence between the phenomenon of accelerating expansion and the anomalous gravitational response. In comparison to the accelerating expansion of the universe \cite{Perlmutter1998,Perlmutter1999,Riess1998,Garnavich1998,Riess2016,Planck2020} which is attributed to the negative pressure from the cosmological constant \cite{Dodelson2008}, our study provides another kind of accelerating expansion, which is the consequence of the interplay among gravity, boundary and quantum effects.

\begin{acknowledgements}
We thank the funding support from the NSFC under Grant Nos.\,12105223, 12247103 and 12175180, Shaanxi Fundamental Science Research Project for Mathematics and Physics under Grant No.\,22JSQ041. J.-H. Zheng also acknowledges the support from the research start-up funding from Northwest University.
\end{acknowledgements}

\end{document}